\documentclass[letter]{aa}
\usepackage{natbib,twoopt}
\bibpunct{(}{)}{;}{a}{}{,} 
\usepackage[varg]{txfonts}
\usepackage{graphicx}

\def\tm{{$^\mathrm{m}$}}
\def\Ks{{K$_\mathrm{s}$}}
\def\MK{{M$_\mathrm{K}$}}
\def\Ms{{M$_{\sun}$}}

\begin{document}
\title{Mortality and dust expulsion in early phases of stellar
  clusters\thanks{Based on observations collected at the European Southern
    Observatory, Chile; program: ESO 82.B-0331}} \subtitle{Evidence from NIR
  photometry of nearby, spiral galaxies}
\titlerunning{Dust expulsion from young stellar clusters}
\author{P.~Grosb{\o}l\inst{\ref{inst1}} \and H.~Dottori\inst{\ref{inst2}}}
\offprints{P.~Grosb{\o}l}
\institute{European Southern Observatory,
  Karl-Schwarzschild-Str.~2, D-85748 Garching, Germany\\
  \email{pgrosbol@eso.org}\label{inst1}
\and
  Instituto de F\'{i}sica, Univ. Federal do Rio Grande do Sul,
  Av. Bento Gon\c{c}alves~9500, 91501-970 Porto Alegre, RS, Brazil\\
  \email{dottori@ufrgs.br}\label{inst2}
}

\date{Received ??? / Accepted ???}

\abstract{It is often argued that young stellar clusters suffer a significant
  infant mortality that is partly related to the expulsion of dust and gas in
  their early phases caused by radiation pressure from hot stars and
  supernovae.  Near-infrared (J-K)--\MK\ diagrams of young stellar clusters in
  nearby spiral galaxies show a bi-modal distribution that is consistent with
  a fast decline of their intrinsic extinction at an early epoch. }
{The distinct features in the color-magnitude diagrams (CMD) and the fast
  change of colors for the youngest clusters allow us to place constraints on
  their early evolutionary phases, including the time scale for the decreasing
  extinction caused for instance by gas and dust expulsion.}
{Monte Carlo simulations of cluster populations were performed using the
  power-law distribution function $g(M, \tau) \propto M^{\alpha}
  \tau^{\gamma}$.  Integrated colors were computed from Starburst99 models.
  The simulated near-infrared CMD were compared with those observed for six
  grand-design, spiral galaxies using statistical goodness-of-fit tests. }
{The CMDs indicate a significant mortality of young, massive clusters with
  $\gamma$ = -1.4$\pm0.5$. High initial extinction A$_{V}$ = 8-11\tm\ and
  strong nebular emission are required to reproduce the bi-modal color
  distributions of the clusters.  An extended star formation phase of longer
  than 5\,Myr is suggested.  The reduction of the internal extinction of the
  clusters starts during their active star formation and lasts for a period of
  5-10\,Myr. }{}
\keywords{Stars: formation - Galaxies: star clusters: general - 
  Galaxies: spiral - Methods: statistical - Infrared: general}
\maketitle
\section{Introduction}
Based on the statistics of embedded clusters in the Galaxy, \citet{lada03}
concluded that a high fraction of young proto-clusters will dissolve while
only a few percent are likely to become bound clusters.  Several reasons for
this `infant mortality' have been brought forward, such as an expulsion of
dust and gas by hot stars or supernovae \citep{tutukov78, whitworth79,
  goodwin06, bastian06} and dynamic effects \citep{lamers06, gieles07,
  wielen77}.  The main effect of the dust and gas expulsion is the reduction
of the total cluster potential, which makes the cluster more vulnerable to
perturbations and eventual disruption.

The dissolution time for clusters has been evaluated by comparing the sizes of
old and young cluster populations assuming a nearly constant formation rate.
This was performed for the solar neighborhood \citep{lamers05, piskunov06},
whose dissolution time $t_4^\mathrm{dis}$ was estimated to be in the range of
0.3-1.0\,Gyr for a 10$^4$\,\Ms\ cluster with a power-law dependency on mass.
Several nearby galaxies were investigated \citep{boutloukos03, scheepmaker09}
and show a significant spread in $t_4^\mathrm{dis}$ from 40\,Myr for the inner
parts of M51 to 8\,Gyr for the Small Magellanic Cloud.  \citet{chandar10} used
a cluster distribution function (CDF) $g(M, \tau) \propto M^{\alpha}
\tau^{\gamma}$ depending on cluster mass $M$ and time $\tau$ to analyze the
clusters in the Large Magellanic Cloud. They determined an age exponent
$\gamma = -0.8$, while \citet{gieles08} found a flat distribution with no
significant age dependency.

The distributions of cluster complexes in ten nearby, grand-design spiral
galaxies were studied by \citet{grosbol12} using near-infrared (NIR) colors.
They found observational evidence for a fast reduction of the extinction in
young clusters because their colors were clearly separated from those of the
older clusters with lower reddening.

In the current paper, we analyze the NIR color-magnitude distributions of
cluster complexes in several nearby spirals to estimate parameters for the
dust expulsion phase and the dissolution of young clusters.  The data and
models used to reproduce them are described in the two following sections.
The fitting procedure and general behavior of the main model parameters are
given in section~\ref{sec:behavior}, while results and conclusions are
provided in the last section.

\section{Data}
\label{sec:data}
We selected six of the grand-design spirals of the study by \citet{grosbol12},
for which more than 2000 cluster complexes were identified.  The galaxies were
observed in the NIR JH\Ks-bands with HAWK-I at the Very Large Telescope and
are listed in Table~\ref{tbl:galaxies} together with their assumed distances
estimated from their systemic velocity relative to the 3K cosmic microwave
background using a Hubble constant of 73 km~s$^{-1}$~Mpc$^{-1}$.  The average
seeing on the \Ks-maps was around 0\farcs4, which yields a linear resolution
in the range of 20-40~pc.  This is not enough to distinguish individual
clusters, therefore many of the sources detected are likely to be cluster
complexes.  The total number of sources N$_s$ detected on the \Ks-images and
the limiting magnitude K$^l$ for a 90\% completeness level are provided in
Table~\ref{tbl:galaxies}.

\begin{table}
  \caption[]{List of galaxies. Name, adopted distance D in Mpc, and limiting
    magnitudes K$^l$ for a 90\% completeness level are listed.  The total
    number of sources N$_s$ for which aperture photometry could be obtained is
    given as well as the absolute magnitude limit M$_K^l$ and the
    corresponding number of non-stellar objects N$_c$.  Finally, the CDF
    exponent $\alpha$ derived for young clusters is listed. }
 \label{tbl:galaxies}
 \begin{tabular}{lrrrrrr} \hline\hline
  Galaxy &  \multicolumn{1}{c}{D} &
  \multicolumn{1}{c}{K$^l$}  & \multicolumn{1}{c}{N$_s$} &
  \multicolumn{1}{c}{M$_K^l$} & \multicolumn{1}{c}{N$_c$} & 
  \multicolumn{1}{c}{$\alpha$} \\ \hline 
    \object{NGC\,157}  & 18.0 & 20.2 & 2254 & -11.1 &  569 & -1.62 \\
    \object{NGC\,1232} & 19.8 & 20.6 & 3177 & -10.9 &  927 & -2.37 \\
    \object{NGC\,1365} & 21.1 & 20.2 & 2417 & -11.5 &  827 & -1.97 \\
    \object{NGC\,2997} & 19.2 & 20.1 & 5313 & -11.3 & 1757 & -2.29 \\
    \object{NGC\,5247} & 22.6 & 19.8 & 2259 & -12.0 &  785 & -2.19 \\
    \object{NGC\,7424} &  9.5 & 20.8 & 6137 &  -9.7 & 1212 & -1.76 \\
    \hline
\end{tabular}
\end{table}

\begin{figure}
  \resizebox{\hsize}{!}{\includegraphics{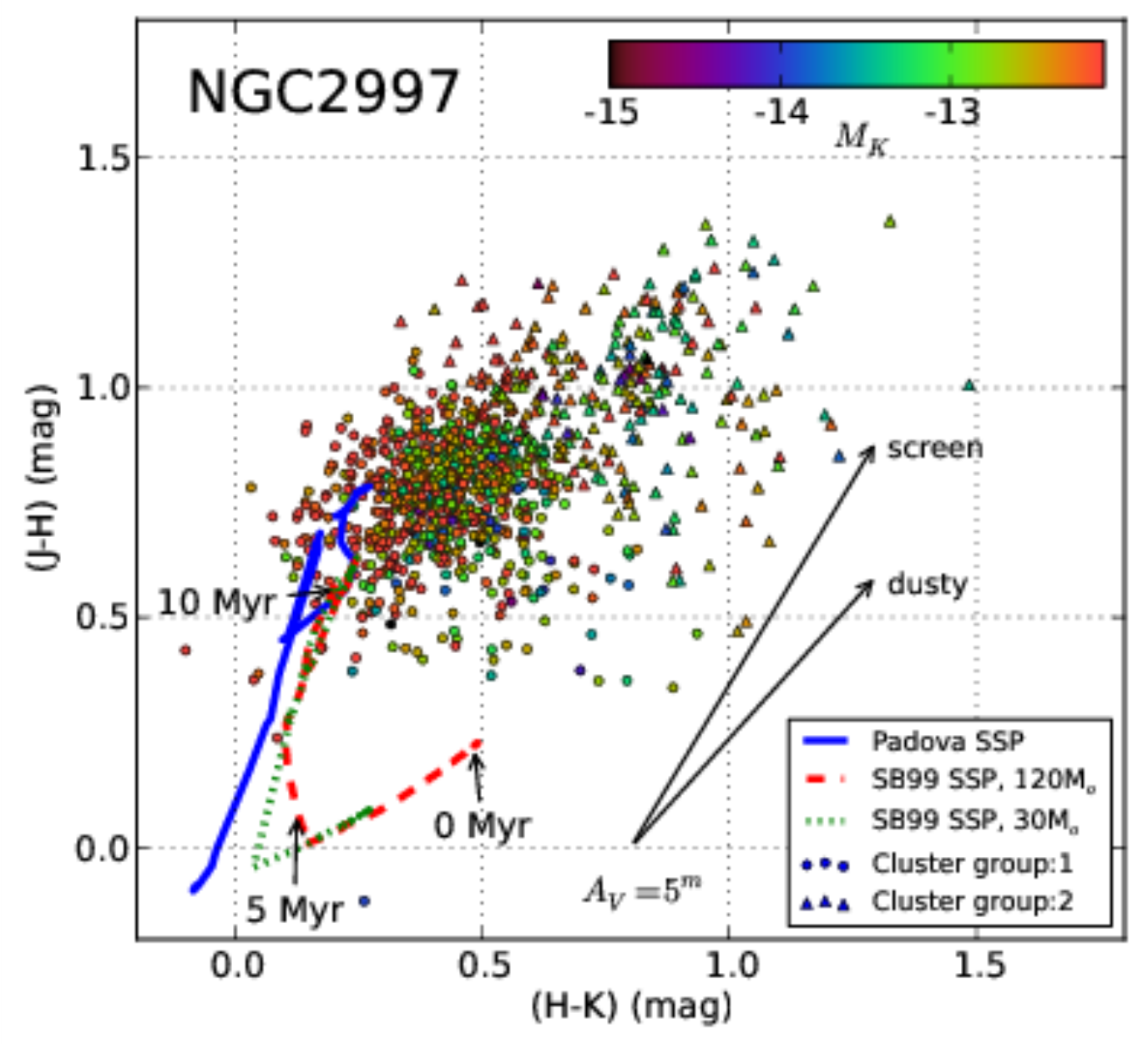}}
  \caption[]{(H-K)--(J-H) diagram of non-stellar sources in NGC\,2997 with
    photometric errors $<$0\fm05.  The magnitudes are indicated by color from
    blue (brighter) to red.  Cluster evolutionary tracks are drawn for the
    Padova and SB99 SSP models. Reddening vectors for screen and dusty models
    are also shown. }
  \label{fig:cc2997}
\end{figure}

\begin{figure}
  \resizebox{\hsize}{!}{\includegraphics{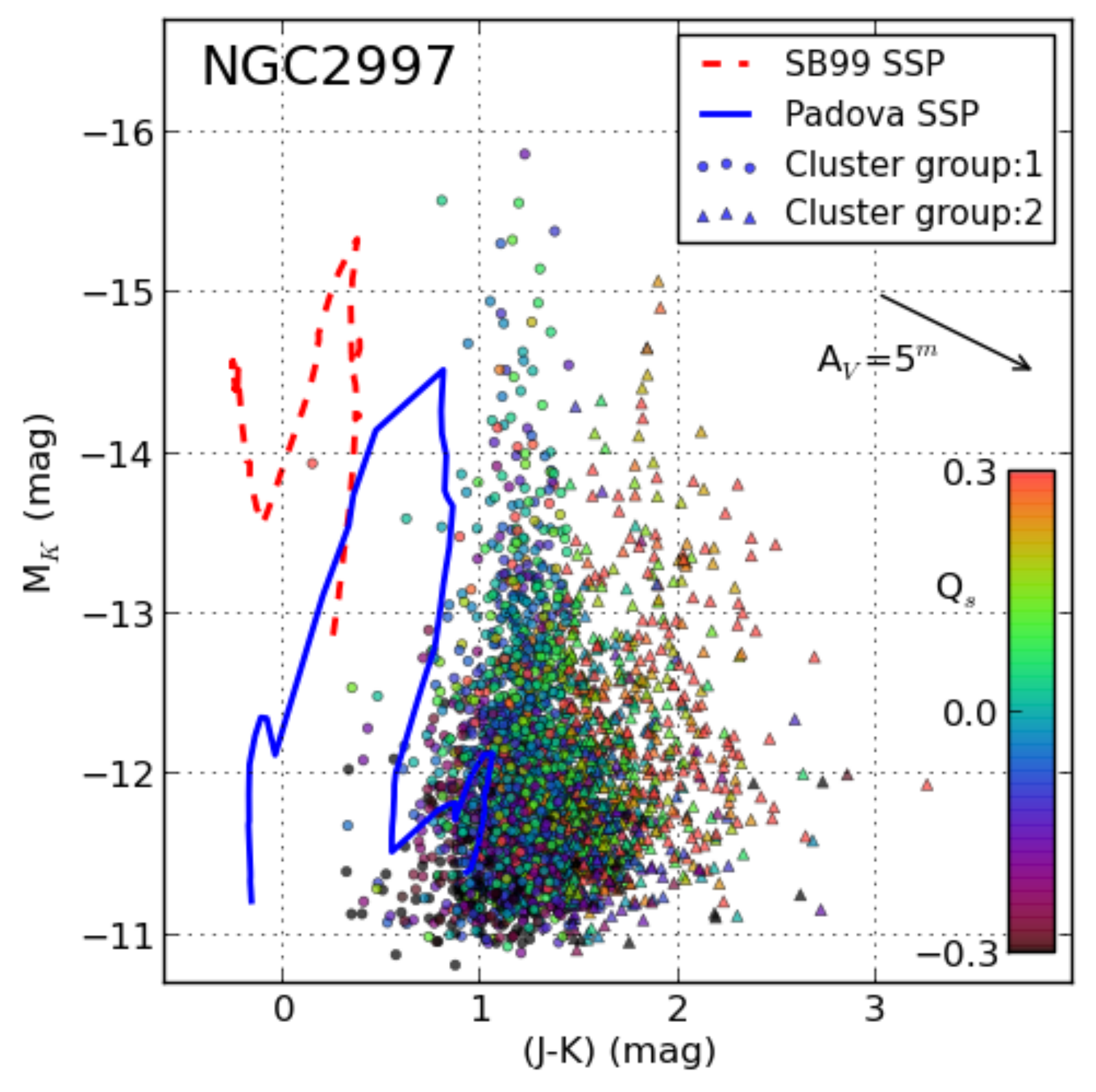}}
  \caption[]{(J-K)--\MK\ diagram for non-stellar objects with photometric
    errors $<$0\fm1 in NGC\,2997.  Evolutionary tracks for clusters with a mass
    of $5\times10^5$\,\Ms\ and a 'screen' reddening vector are shown. The
    symbols indicate the k-means clustering number while colors display the
    reddening-corrected color index Q$_s$ = (H-K) - 0.564$\times$(J-H).}
  \label{fig:cm2997}
\end{figure}

A typical distribution of clusters in a color-color diagram (CCD) is shown for
NGC\,2997 in Fig.~\ref{fig:cc2997}, where cluster evolutionary tracks (CET)
for single-burst stellar population (SSP) models from Padova \citep{marigo08}
and Starburst99 \citep[ hereafter SB99]{leitherer99, vazques05} are plotted
for reference.  The main difference between the two sets of CETs is the
inclusion of nebular emissions in the SB99 models.  Reddening vectors for a
visual extinction $A_V$ = 5\tm\ are also indicated for a standard Galactic
`screen' model \citep{indebetouw05} and a 'dusty' environment \citep{witt92,
  israel98}.  Two groups can be distinguished, a densely populated group close
to the old end of the CETs and one around (0\fm8, 1\fm1), which is composed of
young clusters, strongly attenuated by dust.  These groups can also be seen on
a color-magnitude diagram (CMD) as two separate branches (see
Fig.~\ref{fig:cm2997}), which suggests that clusters suffer a fast reduction
of extinction at an early evolutionary phase.

These two distinct groups of cluster complexes can be identified in all six
spirals.  Whereas their absolute colors will depend on the detailed parameters
of the underlying stellar population (e.g., initial mass function (IMF), and
metallicity), the relative positions of the two groups are more directly
determined by their early history, such as amount of extinction and
evolutionary time scale.  The clusters were separated into two groups by
applying a k-means clustering algorithm \citep{macqueen67} to ensure an
objective procedure.  The grouping is shown in Figs.~\ref{fig:cc2997} and
\ref{fig:cm2997} with different symbols.  The older group contains 60-70\% of
the clusters and their color difference is $\Delta$(J-K) $\approx$ 0\fm7.

\section{Models}
\label{sec:models}
A simple model was created to fit the NIR CMDs of stellar clusters as observed
in the galaxies.  The foundation of the model was a CDF $g(M, \tau) \propto
M^{\alpha} \tau^{\gamma}$ with a power-law dependence on cluster final mass
$M$ and age $\tau$ \citep{chandar10}.  A constant formation rate was assumed
for the galaxies.  The actual mass $M_a$ of a single cluster was computed
assuming that it had a star formation rate SFR = $M/\tau_s$ till $\tau_s$,
after which star formation terminated.  The intrinsic NIR magnitudes of the
clusters were obtained using their actual masses and ages to interpolate in
the SSP models from SB99 (v6.0.2).  The age spread within the youngest
clusters was taken into account by applying appropriate weights.  A Kroupa IMF
\citep{kroupa01} was used for the individual clusters.  The upper limits of
stellar masses $M_u$ for the IMF was varied in the range 30-120\,\Ms.  To
estimate the importance of nebular emission for the intrinsic colors for young
clusters, models with and without emission were computed.

The general distribution in the CCDs indicates that attenuation by dust is
very important at early stages, while later it becomes small.  This was modeled
by applying an extinction $A_V^x$ up to an age $\tau_x$ after which it
decreased linearly to $A_V^o$ over a time $d\tau$.  A simple scenario, where
supernovae expel gas and dust from the cluster and thereby stop star
formation, suggests that $\tau_x\le\tau_s<\tau_x+d\tau$.  In the more general
case where hot stars also erode nearby dust and molecular clouds
\citep{whitworth79}, $\tau_x$ and $\tau_s$ may be similar or even reversed.

The wavelength dependency of extinction was assumed to follow a power-law
$A_V(\lambda) \propto \lambda^{-\beta}$ \citep{martin90}.  A Galactic screen
model for extinction \citep{indebetouw05} is reproduced by $\beta \sim 1.8$,
while 1.3 gives R$_\mathrm{V}$ = 3.0 \citep{turner89}.  Exponents around 0.5
yield values of E(H-K)/E(J-H) typical for integrated light from a dusty,
star-forming environment \citep{witt92, israel98}.  Finally, Gaussian errors
were added to intrinsic colors and extinction applied to simulate
observational errors and spread in cluster initial conditions.

The model code was written in Python using the {\it scipy} package to generate
and analyze the CDFs.  The clusters were created in the age range
$1$-$10^3$\,Myr to cover the observed CMDs.  Their final masses were in the
range of $10^4$-$10^7$\Ms\ except for NGC\,7424, for which a ten times lower
mass range was used to match its fainter clusters.  The model CDF constituted
the empirical probability function to which the observed CDFs were compared
with a Kolmogorov-Smirnov (KS) test. A typical model contained $10^6$ clusters
to ensure that all parts of the distributions were well populated.  This
number was increased, if needed, so that the number of simulated clusters
brighter than the \MK-limit applied was at least five times larger than the
observed population to ensure a smaller statistical fluctuation in the
reference distribution.

\begin{table*}
 \caption[]{Parameters for the best models.  The visual extinctions $A_V^o$
   and $A_V^x$ are given in magnitudes, the times $\tau_s$, $\tau_x$, and
   $d\tau$ are given in log(yr). }
 \label{tbl:res1}
 \begin{tabular}{lrrrrrrrrrr} \hline\hline
   & & \multicolumn{5}{c}{CMD fit} & \multicolumn{4}{c}{CCD fit} \\
   Galaxy  & \multicolumn{1}{c}{M$_u$} &
   \multicolumn{1}{c}{$\gamma$} & \multicolumn{1}{c}{$A_V^o$} &
   \multicolumn{1}{c}{$A_V^x$}  & \multicolumn{1}{c}{$\tau_s$} &
   \multicolumn{1}{c}{\^{D}$_{cmd}$} & \multicolumn{1}{c}{$\beta$}  &
   \multicolumn{1}{c}{$\tau_x$} & \multicolumn{1}{c}{$d\tau$} &
   \multicolumn{1}{c}{\^{D}$_{ccd}$} \\ \hline
 NGC~157 &  60 & -1.1 & 3.4 & 10.6 & 7.0 & 1.11 & 2.1 & 7.1 & 6.4 & 2.08 \\
 NGC~157 & 100 & -1.2 & 3.1 &  8.6 & 6.9 & 1.09 & 1.7 & 6.7 & 6.8 & 1.66 \\
 NGC~157 & 120 & -1.0 & 3.5 &  8.7 & 6.8 & 0.67 & 1.7 & 6.5 & 6.7 & 1.51 \\
\hline
NGC~1232 &  60 & -1.9 & 3.0 & 10.3 & 7.0 & 1.68 & 2.2 & 7.2 & 6.1 & 2.40 \\
NGC~1232 & 100 & -1.9 & 3.3 & 11.6 & 6.9 & 1.12 & 2.5 & 6.3 & 7.1 & 1.86 \\
NGC~1232 & 120 & -1.7 & 3.4 & 10.4 & 6.9 & 1.44 & 2.2 & 6.2 & 7.0 & 1.79 \\
\hline
NGC~1365 &  60 & -1.6 & 3.9 & 10.9 & 6.9 & 1.35 & 2.3 & 7.0 & 6.1 & 2.26 \\
NGC~1365 & 100 & -1.5 & 4.0 & 10.4 & 6.7 & 0.79 & 2.4 & 6.1 & 6.5 & 1.76 \\
NGC~1365 & 120 & -1.1 & 3.8 & 10.2 & 6.7 & 0.65 & 2.2 & 6.1 & 6.6 & 1.80 \\
\hline
NGC~2997 &  60 & -2.4 & 4.1 &  9.2 & 6.8 & 0.86 & 1.6 & 6.0 & 6.9 & 2.21 \\
NGC~2997 & 100 & -1.9 & 3.8 & 10.1 & 6.7 & 0.70 & 1.5 & 6.3 & 6.4 & 2.38 \\
NGC~2997 & 120 & -1.0 & 4.0 & 11.0 & 6.7 & 0.92 & 2.0 & 6.1 & 7.0 & 2.96 \\
\hline
NGC~5247 &  60 & -1.5 & 3.9 & 11.3 & 7.5 & 2.13 & 1.8 & 7.4 & 6.8 & 2.39 \\
NGC~5247 & 100 & -1.5 & 4.2 & 11.7 & 7.0 & 1.16 & 2.2 & 6.9 & 7.0 & 1.63 \\
NGC~5247 & 120 & -1.8 & 3.0 & 11.2 & 7.0 & 1.09 & 1.9 & 6.0 & 7.2 & 1.34 \\
\hline
NGC~7424 &  60 & -1.2 & 3.3 & 11.1 & 7.0 & 1.65 & 2.1 & 7.0 & 6.5 & 3.66 \\
NGC~7424 & 100 & -0.9 & 3.5 & 11.3 & 6.7 & 0.76 & 2.0 & 6.2 & 6.8 & 2.85 \\
NGC~7424 & 120 & -0.9 & 3.9 & 10.3 & 6.6 & 2.27 & 2.2 & 6.4 & 6.5 & 2.77 \\
\hline
\end{tabular}
\end{table*}

\begin{figure}
  \resizebox{\hsize}{!}{\includegraphics{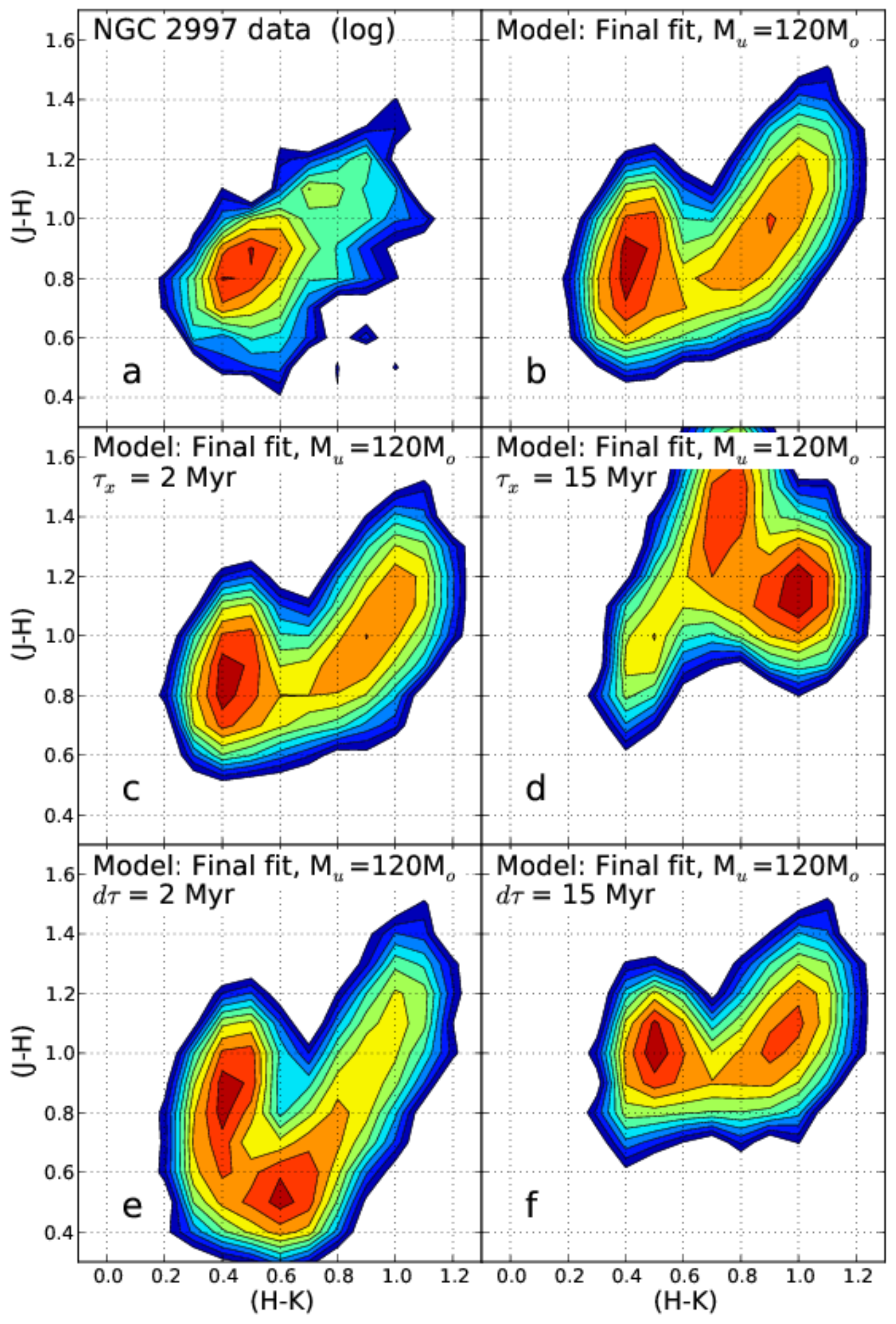}}
  \caption[]{Logarithmic contours of the color distribution of stellar
    clusters in NGC\,2997.  a) Observed cluster distribution, b) model with a
    KS minimum for M$_u$ = 120\,\Ms, while c,d,e, and f show a variation of
    the parameters $\tau_x$ and $d\tau$.  }
  \label{fig:model}
\end{figure}

\section{Fitting procedure and results}
\label{sec:behavior}
The JHK colors of the model clusters depend in a complex way on the eight model
parameters and the adopted CET.  A partial fitting procedure to minimize the
KS test statistics was applied for each CET because different features of the
distributions were more sensitive to some parameters than to others.  First,
the extinctions $A_V^o$ and $A_V^x$ were estimated by fitting the (J-K) values
of the old and young branches in the CMD while their population ratio gave a
preliminary value of $\tau_s$. The magnitude distribution and the relative
importance of the two branches in the CMD are mainly depending on $\alpha$,
$\gamma$, $A_V^x$, $A_V^o$, and $\tau_s$.  Due to the limited age information
available for the old clusters, $\alpha$ and $\gamma$ could not be separated.
Thus, $\alpha$ was fixed to the value derived for young clusters by
\citet{grosbol12} as listed in Table~\ref{tbl:galaxies}.  Only clusters one
magnitude brighter than M$_K^l$ were considered for the fits to avoid any bias
due to incompleteness of faint clusters.  The relative colors of the two
groups are sensitive to the parameters $\beta$, $\tau_x$, and $d\tau$, which
were estimated from the CCD.

The lowest value of the KS test statistics was estimated first by following
the steepest gradient using 0\fm1 and 0\fm5 bins for color indexes and
magnitude, respectively.  Because the test function had an uneven surface and
could have several local minima, a grid of the parameter values was computed
around the minimum found by the gradient search to ensure that the deepest
minimum was located.  Although the model with the KS minimum represents the
`best' fit, it depends mainly on the more populated group of older clusters
and may not reproduce the colors of the young clusters accurately.

Models with $\beta<1.0$ or without nebular emission were also computed but
had in all cases significantly higher KS values. The parameter values for the
three best models are listed in Table~\ref{tbl:res1} including the level of
significance expressed as the population-corrected critical value \^{D}, where
1.22 corresponds to a 10\% level.

The effects of statistical fluctuation cannot be entirely neglected even with
samples of $10^6$ clusters because the high-luminosity tail of the distributions
is significant.  The statistical variation was estimated by computing five
identical models with different seeds for the random number generator.  These
tests indicate that the parameters may fluctuate by 10\% due to Monte Carlo
sampling effects.

The cluster distribution in the CCD is illustrated in Fig.~\ref{fig:model}
with logarithmic contours for both the observed cluster population of
NGC\,2997 and the corresponding model for M$_u$ = 120\,\Ms.  Although the
bi-modal distribution is clearly visible, a significant amount of model
clusters with (H-K) $>$ 0\fm9 does not agree with the observations.  A smaller
amount of nebular emission (e.g., due to a clumpy interstellar medium) would
move the young clusters closer to the location observed.  The effects from
changing $\tau_x$ to 2 and 15\,Myr are displayed in Figs.~\ref{fig:model}c-d,
while Figs.~\ref{fig:model}e-f give a similar variation of $d\tau$.  High
values of $\tau_x$ yield an extra peak in the color distribution with (J-H) =
1\fm4, whereas too many clusters with (J-H) $<$ 0\fm6 are observed for low
values of $d\tau$.  The observed color distribution can best be modeled by an
early start of the decrease in cluster extinction.  The decline is likely to
last for about 10\,Myr.  Most models display a bridge between the young and
old groups that curves toward low (J-H) values and reflects the shape of the
CET.  The selection of a better CET requires detailed spectroscopic
information and is beyond the scope of the current paper.

\section{Discussion and conclusion}
\label{sec:conclusions}

The models suggest that nebular emission is significant and must be included
to account for the NIR colors of young clusters.  A reddening law with $\beta$
around 1.8 or slightly above provides better fits than lower exponents,
but a smaller amount of nebular emission assumed for the CETs (e.g. due to
a clumpy medium) would favor lower values of $\beta$.  A high mass limit
M$_u$ in the range of 100-120\,\Ms\ is preferred in most cases. It is mainly
constrained by the color of the young clusters.

The initial value $A_V^x$ of the average extinction for the cluster complexes
lies in the range of 8-11\tm, while the final extinction $A_V^o$ is about
3-4\tm.  The latter value is close to zero if the Padova CETs are used (see
Fig.~\ref{fig:cm2997}).  These values are consistent with the sources being
complexes of young, highly obscured clusters as seen by \citet{chene13}.  The
linear resolution could also play a role, although no clear trend is apparent.

All fits suggest a higher mortality of the young, massive complexes with
$\gamma$ = -1.4$\pm$0.5 than was found by \citet{fall12} with $\gamma$ = -0.8
for a number of different types of galaxies.  This indicates that the
mortality for very massive complexes is higher than for individual clusters;
that is, young complexes may disintegrate into smaller clusters, which are
below the limiting magnitude of the current study.

The duration of the continuous star formation phase $\tau_s$ is at least
5\,Myr for all galaxies.  In general, the extinction starts to decrease
before the star formation ceases, indicated by $\tau_x < \tau_s$.  Simulations
with the extreme values of 2 and 15 Myr for $\tau_x$ and $d\tau$ favor shorter
$\tau_x$ and longer $d\tau$.

This suggests a star formation scenario where high-mass cluster complexes
(i.e., $M > 10^4$\,\Ms) form stars during an extended period of several Myr.
The reduction of the internal extinction starts before the star formation
terminates.  The time scales suggest that the expulsion of dust is initiated
by the first supernovae of heavy stars.  Owing to the large mass of the cluster
complexes, the first supernovae may not be able to disrupt the giant molecular
cloud (GMC) and star formation continues for some time until enough supernovae
have exploded to destroy the GMC.  A fragmentation of the initial GMC into
individual star-forming regions with slightly different evolution time scales
would also yield a simultaneous reduction of absorption and star formation in
the complexes.

\begin{acknowledgements}
We thank an anonymous referee for helpful comments.  HD thanks the Brazilian
Council of Research CNPq, Brazil, for support.
\end{acknowledgements}
\bibliographystyle{aa}
\bibliography{AstronRef}

\end{document}